\documentclass[]{jfm}

\usepackage{graphicx}
\usepackage{newtxtext}
\usepackage{newtxmath}
\usepackage{natbib}
\usepackage{hyperref}
\usepackage{textcomp}
\hypersetup{
    colorlinks = true,
    urlcolor   = blue,
    citecolor  = black,
}

\newcommand{\RomanNumeralCaps}[1]
\linenumbers

\newcommand\Ra{\mbox{\rm{Ra}}}
\newcommand\Ha{\mbox{\rm{Ha}}}
\newcommand\Nu{\mbox{\textrm{Nu}}}
\newcommand\Rm{\mbox{\rm{Rm}}}
\newcommand\Pm{\mbox{\rm{Pm}}}
\newcommand\Ek{\mbox{\rm{Ek}}}

\title{Wall mode dynamics and transition to chaos in magnetoconvection with a vertical magnetic field}

\author{Matthew McCormack\aff{1},
    Andrei Teimurazov\aff{2},
    Olga Shishkina\aff{2}
\and Moritz Linkmann\aff{1}
\corresp{\email{moritz.linkmann@ed.ac.uk}}}

\affiliation{\aff{1}School of Mathematics and Maxwell Institute for Mathematical Sciences, University of Edinburgh, UK\aff{2}Max Planck Institute for Dynamics and Self-Organization, 37077 Göttingen, Germany}

\begin{document}
\maketitle

\begin{abstract}
Quasistatic magnetoconvection of a low Prandtl number fluid ($\Pr=0.025)$ with a vertical magnetic field is considered in a unit aspect ratio box with no-slip boundaries.  At high relative magnetic field strengths, given by the Hartmann number $\Ha$, the onset of convection is known to result from a sidewall instability giving rise to the wall mode regime.  
Here, we carry out 3D direct numerical simulations of unprecedented length to map out the parameter space at $\Ha = 200, 500, 1000$, varying the Rayleigh number ($\Ra$) between $6\times10^5 \lesssim \Ra \lesssim 5\times 10^8$. We track the development of stable equilibria produced by this primary instability, identify bifurcations leading to limit cycles, and eventually to chaotic dynamics. 
At {$\Ha=200$}, the steady wall mode solution undergoes a symmetry-breaking bifurcation producing a state featuring a coexistence between wall modes and a large-scale roll in the centre of the domain which persists to higher $\Ra$.  However, under a stronger magnetic field at $\Ha=1000$, the steady wall mode solution undergoes a Hopf bifurcation producing a limit cycle which further develops to solutions that shadow an orbit homoclinic to a saddle point.  Upon a further increase in $\Ra$, the system undergoes a subsequent symmetry break producing a coexistence between wall modes and a large-scale roll, although the large-scale roll exists only for a small range of $\Ra$, and chaotic dynamics primarily arise due to a mixture of chaotic wall mode dynamics and arrays of cellular structures.
\end{abstract}

\begin{keywords}
Enter online.
\end{keywords}


\section{Introduction}
Convective flows of electrically conducting fluids influenced by magnetic fields are readily found throughout nature and in many industrial processes.  Examples of such naturally occurring flows arise in the study of stellar convection zones and liquid metal planetary cores \citep{Jones2011}, with industrial applications including liquid metal batteries, casting, semi-conductor crystal growth as well as liquid-metal blanket cooling systems for nuclear fusion reactors \citep{Davidson1999}. 
Despite the physical relevance of such systems, our understanding of magnetoconvection has been limited due to numerous experimental and computational difficulties \citep{schumacher2022various}.  Experiments, which often require the use of opaque liquid metals, are unable to use commonly available optical imaging techniques, and numerical studies usually require substantial computational resource.  
To mitigate these challenges, either greatly simplified systems are used or the simulations are evolved only for short times.  The result is that little is known about the often slowly evolving spatiotemporal dynamics in such systems, especially at high magnetic field strengths.

The onset of magnetoconvection with a vertical magnetic field has been studied in a few isolated scenarios through linear stability analysis: in the case of an infinite/periodic plane layer by \cite{chandrasekhar1961hydrodynamic}; and in the case of a semi-infinite sidewall by \citet{busse2008asymptotic}.  Such analysis reveals that the sidewalls in a system are responsible for the onset of convection at high relative magnetic field strengths (typically characterised by the Hartmann number $\Ha$), particularly relevant to industrial processes that often occur in closed vessels.  Indeed, an analytic-numerical hybrid analysis performed by \citet{houchens2002rayleigh} for the special case of a cylindrical geometry showed that sidewalls were responsible for the onset, producing thin convective layers attached to these sidewalls, commonly referred to as \emph{wall modes}.  However, knowledge about nonlinear effects and the later stages of transition to turbulence are still limited.

Wall modes in quasistatic magnetoconvection with a vertical magnetic field were first examined in the full system numerically by \citet{Liu2018} who conducted direct numerical simulations in a box of aspect ratio $\Gamma=4$ up to $\Ha=2000$ with no-slip boundaries, confirming that onset occurred below the linear stability threshold of the infinite plane layer.  Wall modes were seen to have a 3D spatial structure, featuring thin rolls pressed against the sidewalls in an alternating pattern of positive and negative rotation.  Thus, between each set of rolls lay alternating structures with positive and negative vertical velocities which were additionally seen to have protrusions that extended into the bulk of the domain.  At increased levels of thermal driving (typically characterised by the Rayleigh number $\Ra$), more chaotic cellular style regimes were observed, although details of the transition were not characterised.

Although hints of wall modes were present in the experiments of \cite{cioni2000effect}, wall modes were first confirmed experimentally in a cylinder by \cite{zurner2020flow}, who additionally characterised the cellular regime more carefully, with various numbers of cells being observed at different regions of the $(\Ra,\Ha)$ parameter space.  Further simulations in a cylinder were performed by \citet{akhmedagaev2020turbulent} as well as more recently by \cite{xu2023transition} who combined new experimental and numerical results.  From the current data, it appears that the wall mode protrusions grow with increased $\Ra$, extending towards the bulk of the domain exhibiting quasi-steady dynamics.  Once the critical $\Ra$ of the bulk onset is reached, i.e. the onset in the infinite plane layer \citep{chandrasekhar1961hydrodynamic}, the cellular regime ensues producing a more chaotic variation in system observables such as the dimensionless heat transport (Nusselt number $\Nu$).  For $\Ha < 300$ at mostly higher $\Ra$, \citet{zurner2020flow} observed large-scale rolls, i.e. the large-scale circulation (LSC) that is commonly found in classical Rayleigh-B\'enard convection (RBC), and thus it is assumed that at sufficiently large $\Ra$ magnetoconvection will approach more familiar solutions of RBC.

In summary, although some information about the transition to turbulence is known in magnetoconvection, many questions are still open, mostly concerning the various dynamics admissible by this system.  Namely, it is currently not well understood how the transition between steady wall mode solutions and the more chaotic cellular regime takes place or potentially what states exist in between.  In particular, it is relevant to try and characterise the series of bifurcations that occur upon increased $\Ra$ in this system and how heavily this depends on the relative magnetic field strength $\Ha$.  To address this in the present study, we have performed an array of long numerical simulations in an aspect ratio $\Gamma=1$ box, in many instances on the order of 1000 free-fall times, to study this transition and the associated spatiotemporal dynamics.

Wall modes also occur in rotating Rayleigh-B\'{e}nard convection (RRBC). There are similarities between wall modes in magnetoconvection and in RRBC, especially near onset, where both types of wall modes have a similar two-layer structure. 
This is likely attributed to the extensive similarities in the linear theory between the two systems \citep{busse2008asymptotic, herrmann_busse_1993}. 
However, a number of differences exist between the two systems in terms of their secondary instabilities and further nonlinear effects.  In the rotating case, a Hopf bifurcation leads to an azimuthal precession of the wall modes \citep{ecke1992hopf}, which is not seen to occur in magnetoconvection.  Concerning the spatial structure, striations found in the RRBC wall modes at higher supercriticalities, as observed  by \citet{ecke2022connecting}, have not been found in magnetoconvection.  With that being said, some nonlinear effects do have similarities between the two systems such as wall mode protrusions extending into the bulk, sometimes identified as plume-like jets in the rotating case.  It is yet to be seen if somewhat similar dynamics to those presented here could exists in some regimes of RRBC.

\section{Formulation}
We consider a three-dimensional flow of electrically conducting fluid driven by an imposed vertical temperature difference between the top and bottom of the domain, and in the presence of a vertical magnetic field $\boldsymbol{B}=B_0 \boldsymbol{e}_z$.  Under the quasistatic approximation, that is, for magnetic Reynolds number $\Rm = U\ell/\eta \ll 1$ and magnetic Prandtl number $\Pm = \nu/\eta \ll 1$, where $U$ and $\ell$ are characteristic velocity and length scales, $\eta$ is the magnetic diffusivity, and $\nu$ is the kinematic viscosity, and the Oberbeck-Boussinessq approximation, the velocity field $\boldsymbol{u}$ and the temperature field $T$ evolve from suitable initial conditions according to the following equations,
\begin{subequations}
\label{eq:governing}
\begin{alignat}{1}
    \partial_t\boldsymbol{u} + \boldsymbol{u}\bcdot\bnabla\boldsymbol{u}+\bnabla p &= \sqrt{\frac{\Pr}{\Ra}} \big[\nabla^2\boldsymbol{u} + \Ha ^2(\boldsymbol{j}\times\boldsymbol{e}_z)\big] + T\boldsymbol{e}_z, \\
    \partial_t T + \boldsymbol{u}\bcdot\bnabla T &= \frac{1}{\sqrt{\Ra\Pr}}\nabla^2 T, \\
    \bnabla \bcdot \boldsymbol{u} = 0, \quad 
    \boldsymbol{j} = -\bnabla\phi &+ (\boldsymbol{u}\times\boldsymbol{e}_z), \quad
    \nabla^2\phi = \bnabla\bcdot(\boldsymbol{u}\times\boldsymbol{e}_z),
\end{alignat}    
\end{subequations}
where $p$ is the kinematic pressure, $\boldsymbol{j}$ is the electric current density, $\phi$ is the electric field potential, and $\boldsymbol{e}_z$ is the unit vector that points vertically, opposed to gravity.  
Here, variables have been made dimensionless by using the container height $H$, and the free-fall velocity $u_f=(\alpha g H\delta T)^{1/2}$ to construct length, velocity and time scales, while the temperature difference between the bottom and top plates $\delta T = T_+ - T_-$, and the applied magnetic field strength $B_0$ have been used to construct the dimensionless temperature and magnetic field strength respectively.  In turn, the control parameters of the governing equations (equations \ref{eq:governing}) are the Rayleigh number $\Ra = {\alpha g \delta T H^3}/{\kappa \nu}$, the Prandtl number $\Pr={\nu}/{\kappa}$, and the Hartmann number $\Ha= B_0 H ({{\sigma}/{\rho \nu}})^{1/2}$,
where  $\sigma$ is the electrical conductivity, $\rho$ is the mass density, $\alpha$ is the thermal expansion coefficient, $g$ is the acceleration due to gravity, and $\kappa$ is the thermal diffusivity.
No slip boundary conditions are applied to all boundaries, as well as constant temperatures $T_+$ and $T_-$ applied to the bottom $(z=0)$ and top $(z=H)$ plates respectively.  The domain is equipped with adiabatic side walls $\partial T/\partial \boldsymbol{n}=0$, where $\boldsymbol{n}$ is the vector orthogonal to the surface, and electrically insulating boundaries $\partial \phi/\partial \boldsymbol{n}=0$.

These equations have been solved numerically at a range of $\Ra$ and $\Ha$ for a fluid with $\Pr=0.025$ (such as the GaInSn alloy) in a cubic domain of unit aspect ratio, i.e. $\Gamma = L/H = W/H = 1$ where $L$ and $W$ are the length and width of the domain perpendicular to $\boldsymbol{e}_z$ respectively.  This has been done using the direct numerical solver {\sc goldfish} \citep{reiter2022flow}, which has been widely used in previous studies of convective flows.  This version \citep{teimurazov2023unifying} uses a fourth-order finite-volume discretisation on staggered grids, and a third-order Runge-Kutta time marching scheme  which has been extended to simulate magnetoconvective flows using a consistent and conservative scheme \citep{ni2012consistent}.   
A compromise had to be made in terms of the grid resolution due to the high computational cost.  The flows are resolved on non-uniform grids with $220^2$ points in the cross-plane direction and with either 300 or 350 points in the vertical direction, sufficient to resolve the Hartmann, Shercliff and thermal boundary layers, {with at least 5, 15 and 20 grid points within each layer, respectively}. 
The ratio between the largest finite volume cell length and the Kolmogorov scale, {calculated from the mean dissipation rate}, is less than 3 in the worst case, with the vast majority of simulations being less than 2.  Changes in grid resolution were not seen to qualitatively change the observed dynamics.
Most simulations have been initialised from a lower resolution simulation with $\Ha=0$ with the magnetic field strength being progressively increased, although a few of the solutions have been continued from one another.  In this sense, we try here to broadly classify the states that exist in this flow rather than tracking a detailed route to turbulence.

\section{Numerical Results}
\subsection{Comparison to linear stability theory}
\label{sec:linear-stability}
\begin{figure}
  \raggedleft
  {\includegraphics[width = 12cm]{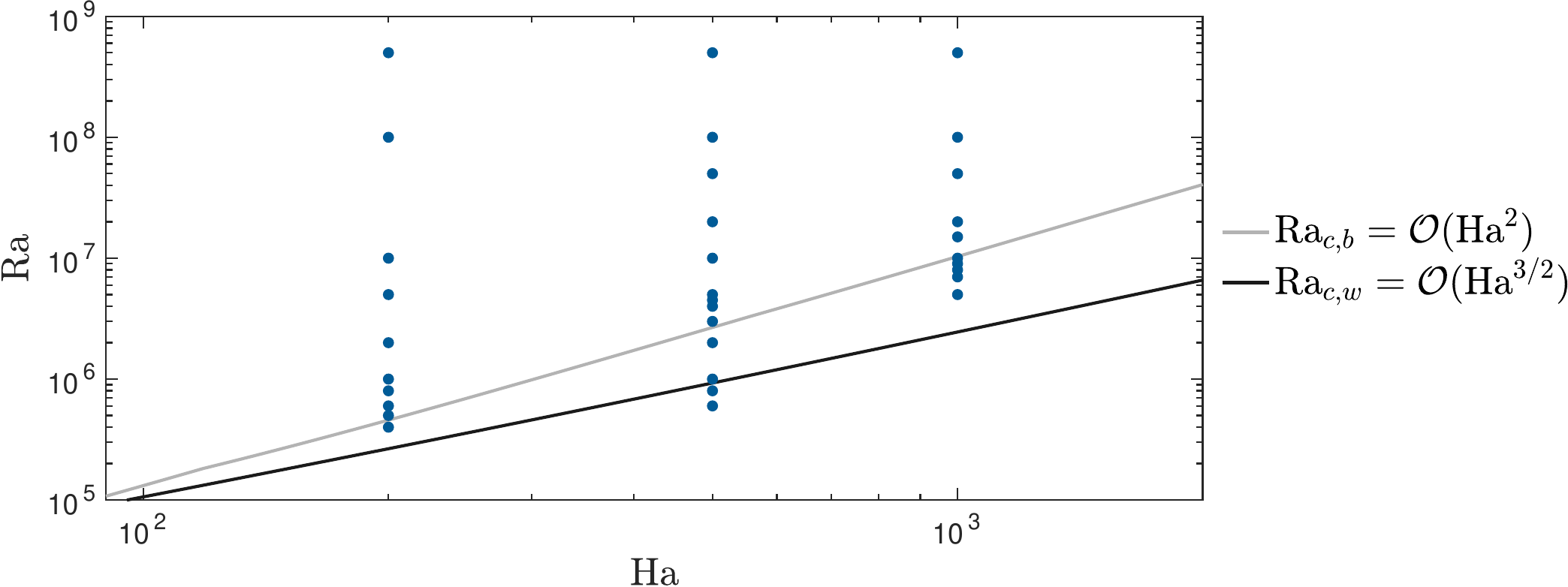}}
  \caption{Studied parameter space given by the Hartmann ($\Ha$) and Rayleigh ($\Ra$) numbers compared to the critical Rayleigh number for onset of the bulk $\Ra_{c,b}$, and wall modes $\Ra_{c,w}$.}
\label{fig:regime-map}
\end{figure}

A number of linear stability results are relevant for the considered system.  Each of these results are derived from perturbations to a static flow field $\boldsymbol{u}=0$, and linear temperature profile.  The dispersion relation for the case of an infinite plane layer/periodic sidewalls and no-slip upper and lower boundaries was derived by \citet{chandrasekhar1961hydrodynamic}.  To apply this result to our cube $(\Gamma=1)$, the dispersion relation is minimised for each value of $\Ha$ over the discrete horizontal wavenumbers admitted by the domain.  This results in the critical Rayleigh number in the bulk $\Ra_{c,b}$ which is plotted in figure \ref{fig:regime-map}.  {At the values of $\Ha$ considered, the wavelength of the most unstable mode in the infinite plane layer is smaller than the size of our domain, and continues to decrease with $\Ha$.  Thus, the bulk onset is seen to be nearly identical in our closed geometry compared to the infinite plane layer.}

Additionally, the influence of no-slip sidewalls is seen to give rise to another linear instability which leads to the formation of thin convective zones close to these sidewalls, known as wall modes \citep{busse2008asymptotic,Liu2018}.  An asymptotic linear stability result to second order, derived by \citet{busse2008asymptotic} for a semi-infinite side-wall bounded domain, gives a relation for the critical Rayleigh number for the side walls denoted by $\Ra_{c,w}$, which has also been plotted in figure \ref{fig:regime-map}.  {The most unstable mode in this case is seen to spatially decay away from the wall with $\Ha^{-1/2}$ at leading order, so becomes increasingly thin compared to our domain with increased magnetic field strength}.  

In both cases, the linear stability results have been derived for the full magnetohydrodynamic (MHD) system.  However, the results are seen in both cases to be independent of the Prandtl number $\Pr$, and the magnetic Prandtl number $\Pm$.  Thus, both results hold in the quasistatic MHD case which we examine here.  The critical Rayleigh numbers for both bulk and wall mode onset have been compared to an overview of our computational data set, shown by the blue markers in figure \ref{fig:regime-map}.  Differences in the velocity boundary conditions are expected to have a small effect on the value of $\Ra_c$ in both cases, with free-slip boundary conditions typically reducing the value of the critical Rayleigh number.

Notably, in the regime covered by our data $(200\leq \Ha \leq 1000)$, the minimum critical Rayleigh number from linear theory is that of the wall mode onset $\Ra_{c,w}$ which is lower than that of the bulk onset, and thus, we expect the onset in our system to occur in the form of a wall mode instability.  At low values of $\Ra$ close to onset we observe equilibrium solutions showing clear evidence of wall modes at each value of $\Ha$.  Our data set at $\Ha=500$ has been extended to below the predicted wall mode onset of \citet{busse2008asymptotic} to as low as $\Ra = 6\times10^5$.  

\begin{figure}
  \centerline{\includegraphics[width=12.5cm]{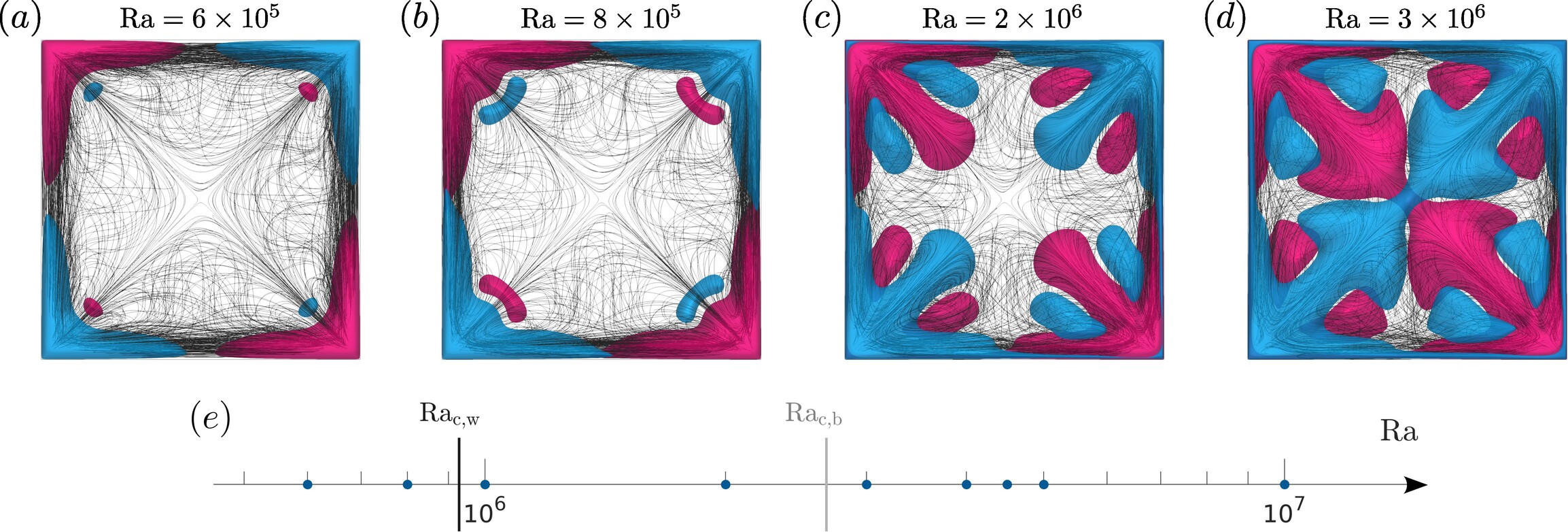}}
  \caption{Overview of the equilibrium solutions at $\textrm{Ha}=500$. $(a$-$d)$ Vertical velocity isosurfaces $u_z = \pm 7\times10^{-3}$ (pink/blue) and instantaneous streamlines (black) from the top view. $(e)$ Comparison of data points to linear theory.}
\label{fig:Ha500-eq}
\end{figure}

\begin{figure}
  \centerline{\includegraphics[width=10.0cm]{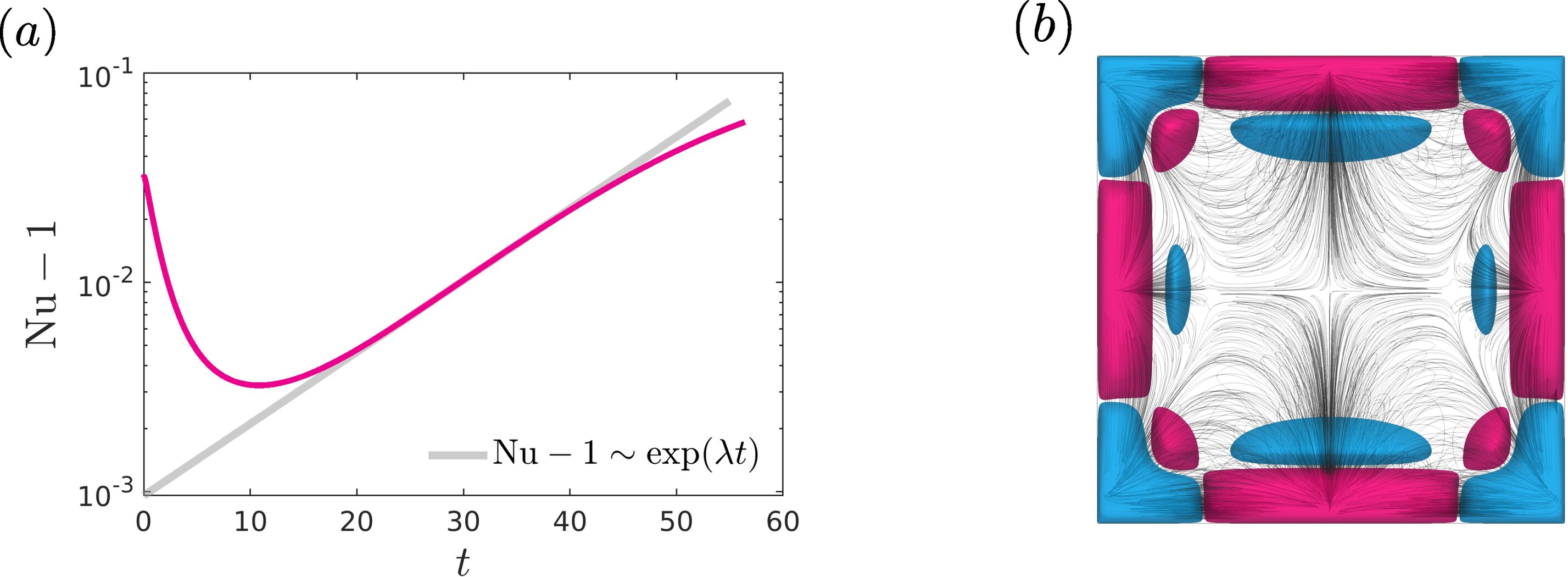}}
  \caption{$(a)$ Temporal evolution of the dimensionless heat transport $\Nu$ from the laminar state perturbed by a cross-flow at $\Ha = 500$, $\Ra=8\times 10^5$.   
  Exponential growth $\Nu -1 \sim \exp(\lambda t)$ is seen, with $\lambda \approx 0.07889$, for a short time before nonlinear growth is observed.  $(b)$ Vertical velocity isosurfaces $u_z = \pm 4\times 10^{-4}$ (pink/blue) and instantaneous streamlines (black) from the top view showing the exponentially growing mode.}
\label{fig:exp-growth}
\end{figure}

Comparing this to the linear stability results (figure \ref{fig:Ha500-eq}$(a,e)$), the onset of the wall mode regime appears to be lower than the asymptotic solution derived in the semi-infinite sidewall case by \citet{busse2008asymptotic} with the convective part of the heat transport being approximately 3\% of the conductive heat transport at $\Ra=6\times10^5$ (i.e. $\Nu = 1.03$).  This indicates that the presence of multiple sidewalls has a significant effect on the critical Rayleigh number, at least at this value of $\Ha$.  Although the onset of convection occurs beneath $\Ra_{c,w}$, the onset here is in line with linear theory, with exponential growth being observed from a perturbed laminar state at $\Ra=8\times10^5$ in figure \ref{fig:exp-growth}$(a)$, and there is no evidence to suggest a subcritical transition. In some instances, we additionally observe equilibrium solutions above the bulk onset $\Ra_{c,b}$, as shown by the equilibrium solution at $\Ra=3\times10^6$ (figure \ref{fig:Ha500-eq}$(d,e)$).  In this case, it is likely that perturbations about the basic state used in the linear stability analysis are not representative of perturbations about a wall-mode dominated flow field and thus the analysis of \cite{chandrasekhar1961hydrodynamic} is not valid in this regime. 

An overview of the equilibrium solutions found at $\Ha=500$ are shown in figure \ref{fig:Ha500-eq}, revealing clear wall mode structures very similar to those found in \citet{Liu2018}.  At $\Ra=6\times10^5$, convective zones are nearly entirely restricted to the near sidewall region, with individual wall modes pinned to each of the four corners from the top view (figure \ref{fig:Ha500-eq}$(a)$) and the velocity isosurfaces being near uniform in the vertical direction.  On each flat sidewall, a roll exists between each wall mode structure, confined to this near-wall region. Small vertical counter-flow structures sit closer to the centre of the domain, in front of each wall mode.  As $\Ra$ is increased, regions of high vertical velocity begin to protrude from the centre of each wall mode, extending into the domain.  This results in a bending of the counter-flow structures at $\Ra=8\times10^5$ (figure \ref{fig:Ha500-eq}$(b)$) subsequently splitting into two distinct counter flow structures by $\Ra=2\times10^6$ (figure \ref{fig:Ha500-eq}$(c)$).  At this point, the wall mode protrusions have extended into the central part of the domain which had near zero velocity at $\Ra=6\times10^5$.  At $\Ra=3\times10^6$ (figure \ref{fig:Ha500-eq}$(d)$), the protrusions have fully extended into the central part of the domain, and show clear interaction with the wall mode in the opposing corner.  

It currently appears that nonlinear effects have a significant influence on the final structure of these wall mode solutions. As can be seen from the time evolution of $\Nu$ from the laminar state perturbed by a cross-flow for $\Ha = 500$ and $\Ra = 8 \times 10^5$ in figure \ref{fig:exp-growth}(a), the Nusselt number grows exponentially after a short transient. Hence, the snapshot shown in figure \ref{fig:exp-growth}(b), taken during this phase of exponential growth, corresponds to the most unstable mode. In comparison with the nonlinear solution shown in figure \ref{fig:Ha500-eq}(b), we observe a slightly different symmetry in the velocity fields. Concerning the heat transport,  we currently observe a deviation from exponential growth at Nu $\approx 1.03$ with the final equilibrium solution having Nu $\approx 1.28$.

Before discussing secondary instabilities, and in view of the aforementioned similarities between wall moded in magnetoconvection and RRBC, we briefly discuss linear stability for RRBC, where more refined analyses have been carried out \citep{liao_zhang_chang_2006, zhang_liao_2009}. Asymptotic analyses for no-slip and stress-free side walls differ at second order in the expansions with the critical Rayleigh number at the onset of near-wall convection is 
$\Ra_w \approx 31.8\Ek^{-1} + 46.49\Ek^{-2/3}$ in the no-slip case, where $\Ek$ is the Ekman number, and $\Ra_w \approx 31.8\Ek^{-1} - 25.25\Ek^{-2/3}$,  in the free-slip case for duct (two rigid walls and one periodic direction) \citep{liao_zhang_chang_2006} and cylindrical geometries \citep{zhang_liao_2009}. In the former case, the asymptotic solutions were compared against a numerical linear stability analysis, assuming two unstable modes, one with retrograde and one with prograde rotation. For the stress-free case, the asymptotics overestimate the critical Rayleigh number for $\Ek \geqslant 10^{-2}$, followed by an underestimation in the range $ 10^{-2}\leqslant \Ek \leqslant 10^{-6}$. A similar trend is observed for the no-slip case, with overestimation for $\Ek \geqslant 10^{-2}$, followed by underestimation in the range $ 10^{-2}\leqslant \Ek \leqslant 10^{-4}$. Apart the asymptotic result reached at lower $\Ek$, the underestimations are less pronounced in the no slip case. 
Further differences may arise for closed rectangular systems.

\subsection{Beyond the primary instability}
We now turn our attention to the dynamics of the flow past the primary wall mode instability.  It is observed that distinct transition processes occur at varying $\Ha$ and thus the strength of the magnetic field plays a pivotal role in the transition to turbulence in this system.

\begin{figure}
  \centerline{\includegraphics[width=12.5cm]{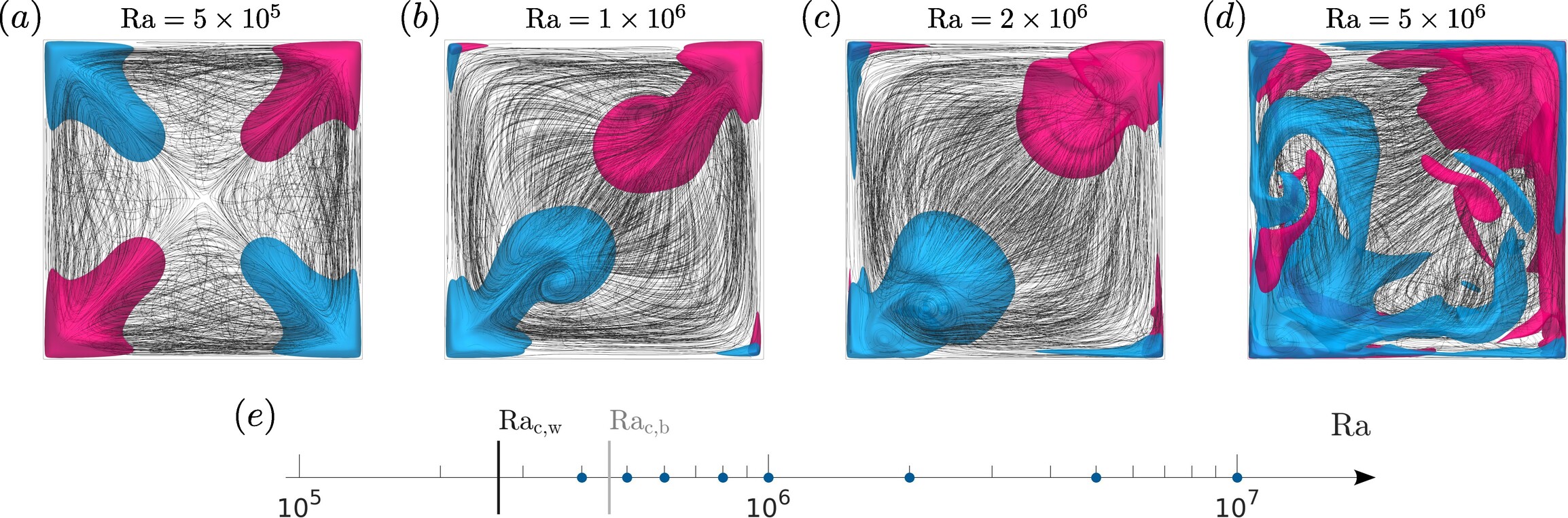}}
  \caption{Overview of the transition at $\textrm{Ha}=200$. $(a$-$h)$ Vertical velocity isosurfaces $u_z = \pm 0.1$ (pink/blue) and instantaneous streamlines (black) from the top view. $(e)$ Comparison of data points to linear theory}
\label{fig:Ha200-overview}
\vspace{1em}
  \centerline{\includegraphics[width=\textwidth]{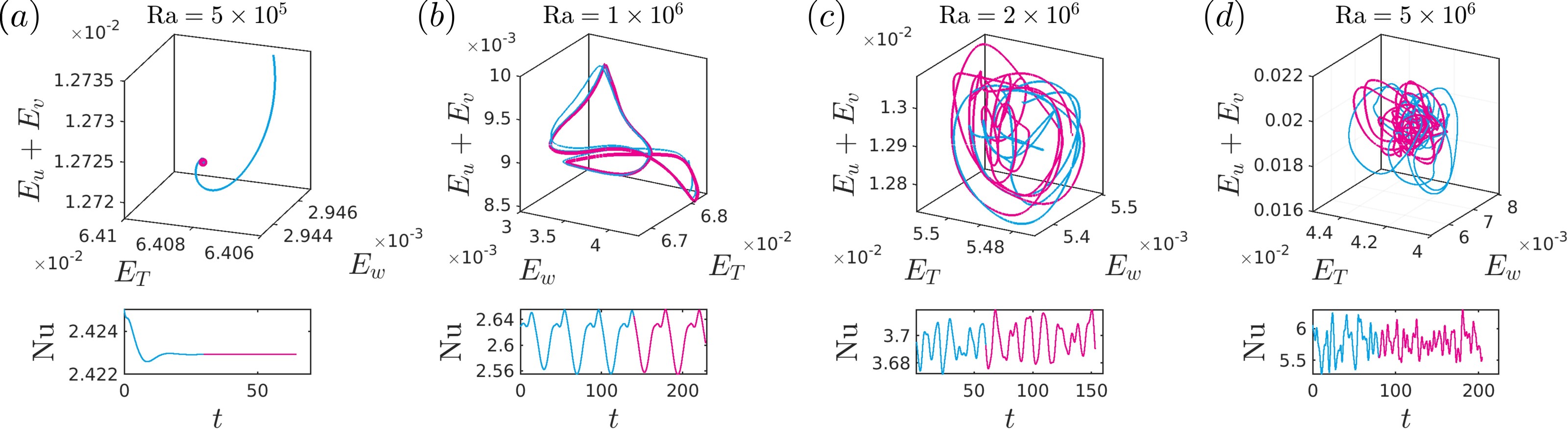}}
  \caption{Phase portrait consisting of thermal energy $E_T$, vertical kinetic energy $E_w$ and cross-plane kinetic energy $E_u+E_v$, and dimensionless heat transport $\Nu$ time series data for each value of $\Ra$ considered at $\Ha=200$.  Here the colours highlight different parts of the $\Nu$ time series for easy comparison with the corresponding phase portrait.}
\label{fig:Ha200-states}
\end{figure}

\begin{figure}
  \centerline{\includegraphics[width=12.5cm]{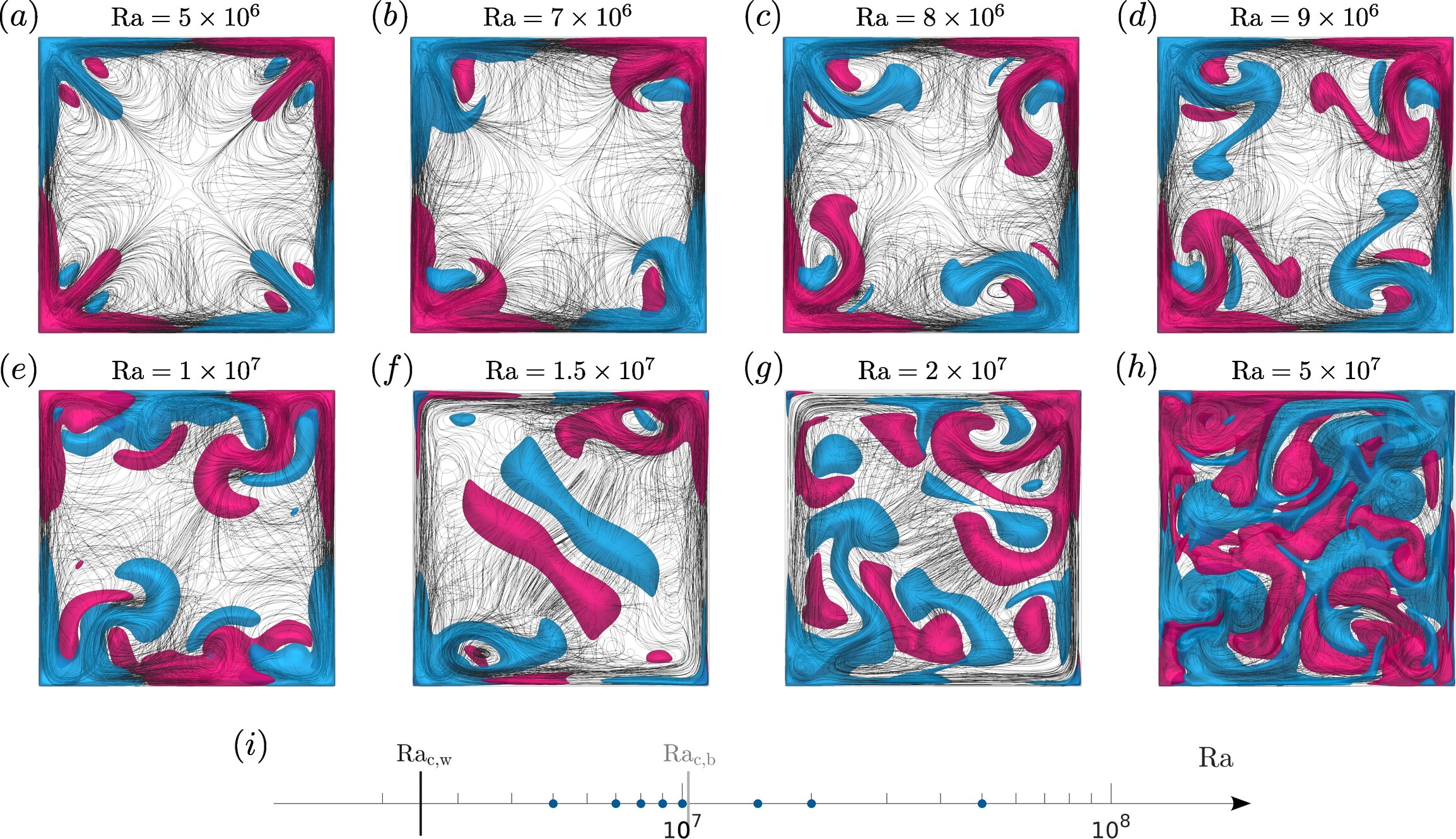}}
  \caption{Overview of the transition at $\textrm{Ha}=1000$. Vertical velocity isosurfaces $(a\textrm{-}e)$ $u_z = \pm 0.01$, $(f\textrm{-}g)$ $u_z = \pm 0.025$ (pink/blue) and instantaneous streamlines (black) from the top view. $(i)$ Comparison of data points to linear theory.}
\label{fig:Ha1000-overview}
\vspace{1em}
  \centerline{\includegraphics[width=\textwidth]{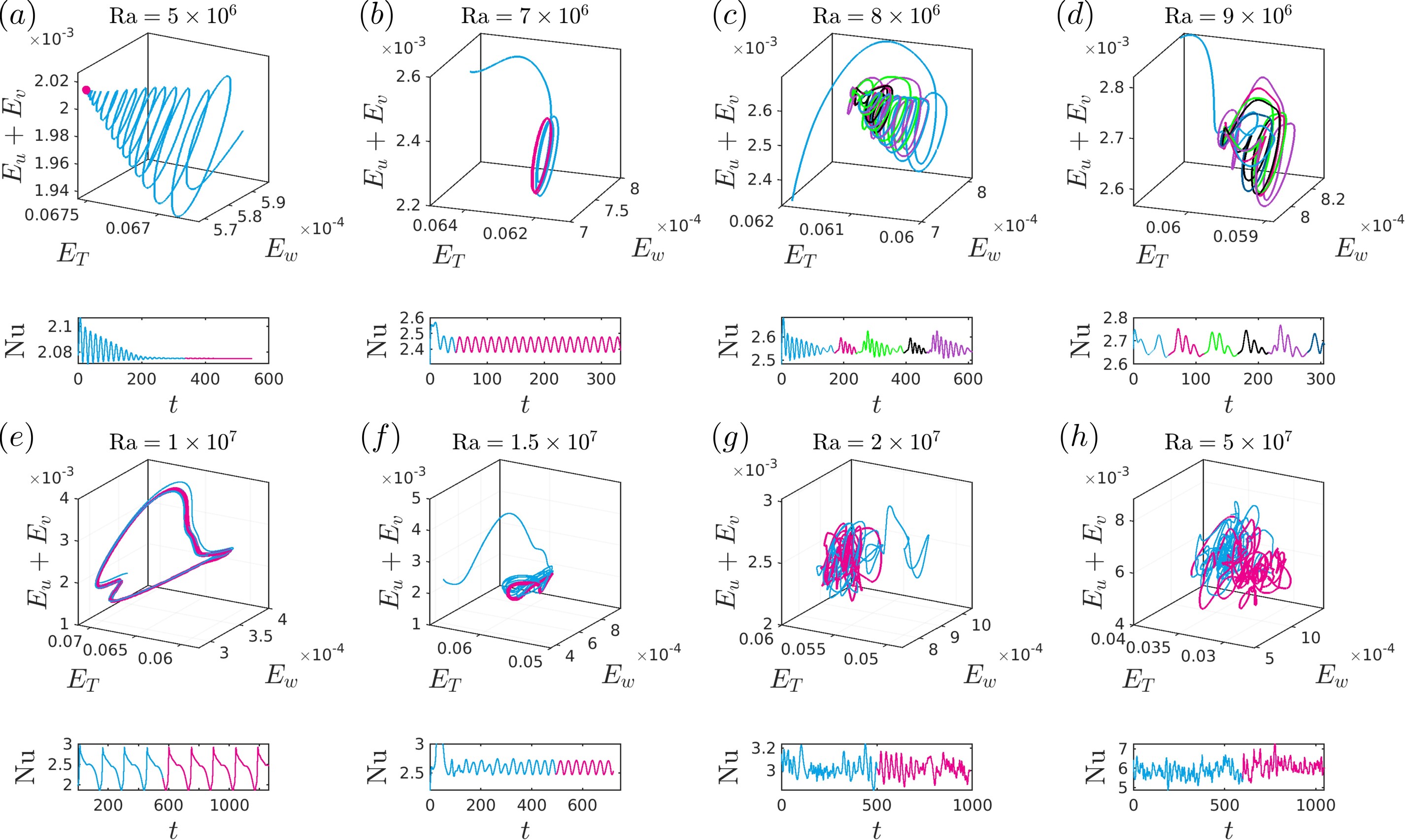}}
  \caption{Phase portrait consisting of thermal energy $E_T$, vertical kinetic energy $E_w$ and cross-plane kinetic energy $E_u+E_v$, and dimensionless heat transport $\Nu$ time series data for each value of $\Ra$ considered at $\Ha=1000$.  Here the colours highlight different parts of the $\Nu$ time series for easy comparison with the corresponding phase portrait.}
\label{fig:Ha1000-states}
\end{figure}

At all values of $\Ha$ considered, the basic wall mode equilibrium solution features a 4-fold symmetry from the top view (90\textdegree ~rotation and flip), and a vertical symmetry about the midplane rotated 180\textdegree ~about $\boldsymbol{e}_z$ (see figures \ref{fig:Ha500-eq}, \ref{fig:Ha200-overview}$(a)$, \ref{fig:Ha1000-overview}$(a)$).  More precisely, the discrete rotational symmetries of the vertical velocity field form a group isomorphic to $\mathbb{Z}_4$ generated by the following symmetry operation,
\begin{equation}
    \mathbb{Z}_4: \: u_z(x,y,z) = -\mathcal{R}_{\pi/2} u_z(x,y,-z),
\end{equation}
which in turn produces the additional symmetry observed $(u_z(x,y,z) = \mathcal{R}_{\pi} u_z(x,y,z))$, where $\mathcal{R}_\beta$ represents a rotation of $\beta$ about $\boldsymbol{e}_z$, and the coordinates $\boldsymbol{x} = (x,y,z)$ have their origin at the centre of the domain.  Another common symmetry observed in the vertical velocity fields is a similar 2-fold symmetry seen in various states whose rotational symmetry forms a group isomorphic to $\mathbb{Z}_2$ generated by the following symmetry operation,
\begin{equation}
    \mathbb{Z}_2: \: u_z(x,y,z) = -\mathcal{R}_\pi u_z(x,y,-z).
\end{equation}

\begin{figure}
  \centerline{\includegraphics[width = 11cm]{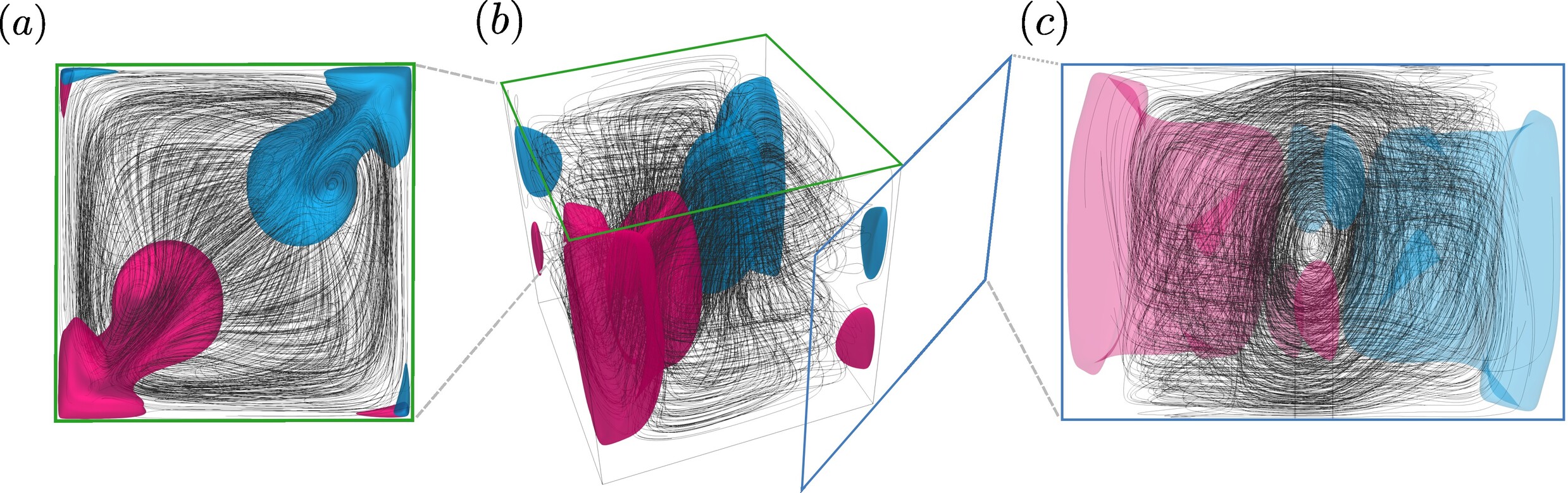}}
  \caption{Vertical velocity isosurfaces $u_z = \pm 0.1$ (pink/blue) and instantaneous streamlines (black) showing coexistence of wall modes and large-scale circulation at $Ha=200$, $Ra=10^6$ from $(a)$ top view, $(b)$ 3d view and $(c)$ angled side view. }
\label{fig:roll_wallmode}
\end{figure}

\begin{figure}
  \centerline{\includegraphics[width=12cm]{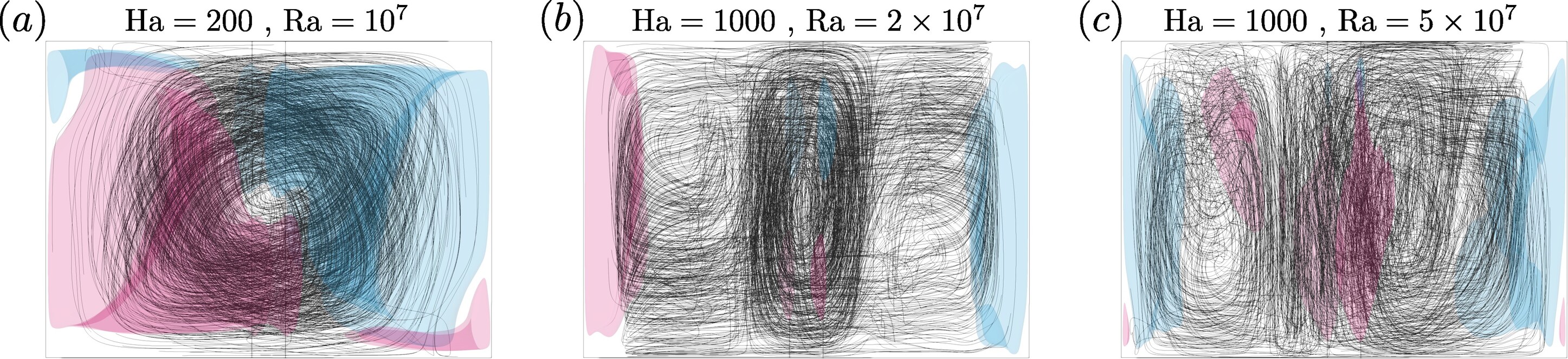}}
  \caption{Mean vertical velocity isosurfaces $(a)$ $u_z = \pm 0.1$, $(b,c)$ $u_z = \pm 0.05$ (pink/blue) and streamlines of mean flow (black) from the angled side view.} 
\label{fig:roll-destruction}
\end{figure}

At $\Ha=200$, wall mode equilibrium solutions are observed at $\Ra=4\times10^5$ to $\Ra=6\times10^5$ (see phase portrait and time series data in figure \ref{fig:Ha200-states}$(a)$) with a $\mathbb{Z}_4$ symmetry (figure \ref{fig:Ha200-overview}$(a)$).  With increased $\Ra$, the wall mode protrusions extend into the centre of the domain, similar to the equilibria at $\Ha=500$ (\S \ref{sec:linear-stability}).  
However, by $\Ra=1\times10^6$, a symmetry-breaking bifurcation has occurred resulting in a state with a $\mathbb{Z}_2$ symmetry (figure \ref{fig:Ha200-overview}$(b)$), which produces a periodic solution (figure \ref{fig:Ha200-states}$(b)$) with a period $\mathcal{T}=82.8$, and marks the breakdown of the rolls near the side walls which existed between the wall modes.  Physically, the wall modes at $\Ra=1\times10^6$ consist of two vortices, one near the corner of the domain and one near the outer tip of the wall mode, the outermost of which is seen to contract and extend away from and into the centre of the domain, periodically switching between being biased to one side of the diagonal and then the other (animated versions of figures \ref{fig:Ha200-overview} and \ref{fig:Ha1000-overview} can be found in the supplementary materials).  Of note, is that wall modes at $\Ra=1\times10^6$ coexist with a large-scale roll in the centre of the domain, orientated about the diagonal as shown in figure \ref{fig:roll_wallmode}.  At these values of $\Ra$, the orientation of the large-scale roll is seen to oscillate back and forth with the wall modes as they contract and expand.  
However, as $\Ra$ is increased to $2\times10^6$, the large-scale roll begins to dominate the dynamics, maintaining a fixed diagonal orientation and pinning the wall modes into the position shown in figure \ref{fig:Ha200-overview}$(c)$.  The wall modes, now pinned close to the corners, produce higher frequency oscillations due to a rapid churning of the wall mode vortex closest to the wall.  The combined effect of these processes is a weakly aperiodic signal shown in figure \ref{fig:Ha200-states}$(c)$.  Upon further increase of $\Ra$, the $\mathbb{Z}_2$ symmetry is seen to break, and aperiodic solutions are observed at $\Ra=5\times10^6$ and $\Ra=1\times10^7$ (figure \ref{fig:Ha200-states}$(d)$).  These solutions exhibit chaotic behaviour in the velocity fields and are seen to contain more small-scale structures (figure \ref{fig:Ha200-overview}$(d)$), which appear to be associated with more chaotic wall mode dynamics that invade the bulk of the domain.  Although masked by the small-scale chaotic dynamics in the instantaneous velocity fields, clear evidence of a large-scale roll can still be seen in the mean flow at these higher Rayleigh numbers (see figure \ref{fig:roll-destruction}$(a)$ for example).

At $\Ha=1000$, the wall mode equilibrium solution appears to have a similar structure to those at lower magnetic field strengths, featuring the same $\mathbb{Z}_4$ symmetry, although the characteristic length scale of the wall modes is affected by the magnetic field.  The result of this is that the wall-mode protrusions become shorter and thinner compared to the length scale along the sidewall (compare figure \ref{fig:Ha1000-overview}$(a)$ to figure \ref{fig:Ha500-eq}$(c)$ for example).  This is consistent with the linear theory which suggests that the most unstable sidewall mode spatially decays into the bulk with $\Ha^{-1/2}$ at leading order, thus producing a thinner sidewall layer at higher $\Ha$ \citep{busse2008asymptotic}.  At increased $\Ra$, the equilibrium solution undergoes a Hopf bifurcation, resulting in the simple limit cycle oscillation ($\mathcal{T}=13.8$) observed at $\Ra=7\times10^6$ (figure \ref{fig:Ha1000-states}$(b)$) which maintains the $\mathbb{Z}_4$ symmetry.  This periodic signal arises due to an oscillation of the wall mode protrusions themselves which synchronously flap back and forth across the diagonal (figure \ref{fig:Ha1000-overview}$(b)$).  As $\Ra$ is further increased to $\Ra=8\times10^6$, the amplitude of the wall mode nose oscillations increases as the protrusions extend further into the domain (figure \ref{fig:Ha1000-overview}$(c)$), appearing to violently crash into the sidewall rolls between the wall modes, which modulates the amplitude of the flapping, damping the wall mode protrusions to a straighter position, closer to the equilibrium solution seen at $\Ra=5\times10^6$.  The protrusions then undergo an instability and quickly return to a large amplitude flapping motion, which damps again, continuing the cycle.  Autocorrelation of the $\Nu$ time series exhibits decays that scale as $\exp({-\lambda\tau})$ with $\lambda \approx 0.041$ for timelag $\tau$. 
The mathematical interpretation of this behaviour is made clearer by the phase portrait and time series data in figure \ref{fig:Ha1000-states}$(c)$ where the solution appears to decay along the stable manifold of a saddle point before being ejected along the unstable manifold of the saddle i.e. the solution appears to shadow an orbit homoclinic to a saddle focus with the saddle point having a similar flow structure to that of the equilibrium found at $\Ra=5\times10^6$.  This behaviour is reminiscent of the Shilnikov phenomenon \citep{wiggins2013global}, arising in the third-order system first studied by \citet{shilnikov1965case}.  By $\Ra=9\times10^6$, the ejections are more frequent (figure \ref{fig:Ha1000-states}$(d)$), with $\Nu$ autocorrelations decaying with $\lambda\approx0.078$, and the protrusions exhibit stronger and more complex flapping behaviour, as seen in figure \ref{fig:Ha1000-overview}$(d)$.  
At $\Ra=1\times10^7$, the flow is observed undergoing a break in symmetry, resulting in a bursting limit cycle with period $\mathcal{T}=147.2$ (figure \ref{fig:Ha1000-states}$(e)$).  The velocity field reveals that three wall mode structures move across opposing sidewalls, which when coinciding with the end sidewall produces a large flap from the end wall mode in opposing corners, as shown in figure \ref{fig:Ha1000-overview}$(e)$, which then decays to the original flow field.  This large flap produces a large increase in the dimensionless heat transport resulting in the bursting behaviour observed in the Nusselt number time series data shown in figure \ref{fig:Ha1000-states}$(e)$.  
At $\Ra=1.5\times10^7$, the $\mathbb{Z}_4$ symmetry has fully broken and has been replaced by the $\mathbb{Z}_2$ symmetry.  This state sustains large periodic wall mode flapping in two opposing corners which converges to the limit cycle ($\mathcal{T}=31.2$) shown in figure \ref{fig:Ha1000-states}$(f)$.  Interestingly, this state also features steady convection in the bulk which occurs at a Rayleigh number higher than $\Ra_{c,b}$ (figure \ref{fig:Ha1000-overview}$(i)$), and features a thin roll, aligned along the diagonal, in the centre of the domain (figure \ref{fig:Ha1000-overview}$(f)$). Additionally, as $\Ra$ is increased further to $2\times10^7$, this large-scale roll appears in the mean flow field (figure \ref{fig:roll-destruction}$(b)$), and the bulk is dominated by vertical flow structures which interact heavily with now chaotically flapping wall mode protrusions which extend far into the domain and shed large-scale structures into the bulk as shown at $\Ra=2\times10^7$ in figure \ref{fig:Ha1000-overview}$(g)$.  However, by $\Ra=5\times10^7$, no evidence of such large-scale roll is observed in the instantaneous or mean flow (figure \ref{fig:roll-destruction}$(c)$) unlike the $\Ha=200$ case.  Chaotic solutions ensue, with a combination of unsteady bulk dynamics in the form of interconnected columnar structures and the remnants of more chaotic wall mode dynamics in the corners similar to the cellular regimes observed previously \citep{Liu2018,zurner2020flow,akhmedagaev2020turbulent,xu2023transition} .  The destruction of the large-scale roll at $\Ha=1000$ has occurred at most by $\Ra/\Ra_{c,b} \approx 4.84$ ($\Ra=5\times 10^7$), whereas at $\Ha=200$, the large-scale roll as been seen to persist in our simulations up to at least $\Ra/\Ra_{c,b} \approx 218.36$ ($\Ra=1\times 10^8$). This shows the clear impact of the magnetic field on the large-scale circulation.

The transition process near onset at $\Ha=500$ is seen to be qualitatively more similar to the transition at $\Ha=1000$, with a Hopf bifurcation leading to a limit cycle with a $\mathbb{Z}_4$ symmetry, which subsequently undergoes a break in symmetry to a $\mathbb{Z}_2$ state.  However, the highest $\Ra$ simulations ($\Ra = 5\times10^8$) carried out at $\Ha=500$ show a large-scale roll in the mean flow like in the $\Ha=200$ case at high supercriticalities, and different from $\Ha=1000$ case where a cellular regime is observed at high $\Ra$.

\section{Conclusions}
We have studied three-dimensional quasistatic magnetoconvection for Hartmann numbers between $200 \leq \Ha \leq 1000$ in a cube $(\Gamma=1)$ with no-slip boundaries, tracking the base state to states exhibiting chaotic multi-scale dynamics.  In line with linear theory in semi-infinite domains, we confirm that the primary instability in this system comes from the sidewalls, giving rise to an equilibrium wall mode solution for all values of $\Ha$ considered.  However, the later stages of the transition to chaos past this primary instability are seen to vary with $\Ha$.  At $\Ha=200$, the basic wall mode state undergoes a symmetry breaking from a $\mathbb{Z}_4$ to a $\mathbb{Z}_2$ symmetry, giving rise to limit cycles involving a large-scale roll which re-orientates in time with wall mode oscillations.  On increased Rayleigh numbers, the large-scale roll is seen to persist accompanied by more chaotic dynamics from the wall modes.  However, at $\Ha=1000$, the equilibrium wall mode state undergoes a Hopf bifurcation resulting in limit cycles involving synchronous oscillations of the wall mode protrusions, which further evolves to states that shadow an orbit homoclinic to a saddle focus involving more complex oscillatory behaviour.  The system then undergoes a $\mathbb{Z}_4$ to $\mathbb{Z}_2$ symmetry breaking bifurcation producing limit cycles featuring a large-scale roll which is dominated by more vigorous wall mode oscillations, which subsequently begin to shed large-scale structures into the bulk.  At higher $\Ra$ this develops to form the cellular regime, and no large-scale roll persists.  

The degree to which the results of this study generalise to other geometries, aspect ratios {and magnetic boundary conditions} is not currently clear.  In larger aspect ratio boxes, it is expected that similar dynamics could occur, and indeed some evidence of potential wall mode nose flapping can be seen in some flow snapshots in the $\Gamma=4$ system of \cite{Liu2018}, although short run times of these simulations mean that this cannot be confirmed, and further, whether such oscillations are synchronised between the larger number of wall modes.  In cylindrical containers there is also some evidence of extended wall mode protrusions and potential dynamics in the presented flow snapshots of \citet{zurner2020flow} and \cite{xu2023transition}, although the continuous symmetry in this geometry appears to result in less strict wall mode symmetries in some instances and dynamics such as those described here have not yet been identified.  It is currently unclear whether wall modes persist in more generalised geometries.  However, wall modes in rotating convection have been seen to persist independently of the geometry in both numerical and experimental studies \citep{favier2020robust,ecke2022connecting}.  With respect to magnetic boundary conditions, we expect that the results here would generalise to low but finitely conducting boundaries, such as those found in experiments. The reason being that wall modes arise due to a suppression of the Lorentz force near the boundary, largely due to electrically insulated boundaries. However, an investigation into the subsequent bifurcations from the basic wall mode state and dynamics of the ensuing states as discussed here has not yet been carried out experimentally.

Here, we focused on the identification of a large number of flow states including equilibria, limit cycles and chaotic dynamics to map out the parameter space and to compare with results from linear stability of wall-mode and bulk onset.  Further analysis is currently underway to describe the transition to turbulence, for instance through a series of bifurcations. Preliminary results suggest the presence of multiple states and hysteresis. In the context of pattern formation, the possibility of hysteresis in this system has been suggested in experimental studies \citep{zurner2020flow}.

\backsection[Supplementary data]{\label{SupMat}Supplementary material and movies are available at ...}

\backsection[Acknowledgements]{The authors thank G. M. Vasil for fruitful discussions.}

\backsection[Funding]{This work was supported by the Deutsche Forschungsgemeinschaft (SPP1881 ”Turbulent Superstructures” grants Sh405/7, Sh405/16 and Li3694/1) and used the ARCHER2 UK National Supercomputing Service (https://www.archer2.ac.uk) with resources provided by the UK Turbulence Consortium (EPSRC grants EP/R029326/1 and EP/X035484/1).}

\backsection[Declaration of interests]{The authors report no conflict of interest.}
 
\bibliographystyle{jfm}
\bibliography{jfm}

\begin{thebibliography}{22}
\expandafter\ifx\csname natexlab\endcsname\relax\def\natexlab#1{#1}\fi
\def\au#1{#1} \def\ed#1{#1} \def\yr#1{#1}\def\at#1{#1}\def\jt#1{\textit{#1}} \def\bt#1{#1}\def\bvol#1{\textbf{#1}} \def\vol#1{#1} \def\pg#1{#1} \def\publ#1{#1}\def\arxiv#1{#1}\def\org#1{#1}\def\st#1{\textit{#1}}

\bibitem[Akhmedagaev {\em et~al.\/}(2020)Akhmedagaev, Zikanov, Krasnov \& Schumacher]{akhmedagaev2020turbulent}
{\sc \au{Akhmedagaev, R.}, \au{Zikanov, O.}, \au{Krasnov, D.} \& \au{Schumacher, J.}} \yr{2020}  \at{Turbulent {Rayleigh--B{\'e}nard} convection in a strong vertical magnetic field}.  \jt{J. Fluid Mech.}  \bvol{895},  \pg{R4}.

\bibitem[Busse(2008)]{busse2008asymptotic}
{\sc \au{Busse, F.~H.}} \yr{2008}  \at{Asymptotic theory of wall-attached convection in a horizontal fluid layer with a vertical magnetic field}.  \jt{Phy. Fluids}  \bvol{20}~(2).

\bibitem[Chandrasekhar(1961)]{chandrasekhar1961hydrodynamic}
{\sc \au{Chandrasekhar, S.}} \yr{1961}  \at{Hydrodynamic and hydromagnetic stability}.  \jt{Int. Ser. Monogr. on Phys.} .

\bibitem[Cioni {\em et~al.\/}(2000)Cioni, Chaumat \& Sommeria]{cioni2000effect}
{\sc \au{Cioni, S.}, \au{Chaumat, S.} \& \au{Sommeria, J.}} \yr{2000}  \at{Effect of a vertical magnetic field on turbulent {Rayleigh-B{\'e}nard} convection}.  \jt{Phys. Rev. E}  \bvol{62}~(4),  \pg{R4520}.

\bibitem[Davidson(1999)]{Davidson1999}
{\sc \au{Davidson, P.~A.}} \yr{1999}  \at{{Magnetohydrodynamics in materials processing}}.  \jt{Annu. Rev. Fluid Mech.}  \bvol{31},  \pg{273--300}.

\bibitem[Ecke {\em et~al.\/}(2022)Ecke, Zhang \& Shishkina]{ecke2022connecting}
{\sc \au{Ecke, R.~E.}, \au{Zhang, X.} \& \au{Shishkina, O.}} \yr{2022}  \at{Connecting wall modes and boundary zonal flows in rotating rayleigh-b{\'e}nard convection}.  \jt{Phys. Rev. Fluids}  \bvol{7}~(1),  \pg{L011501}.

\bibitem[Ecke {\em et~al.\/}(1992)Ecke, Zhong \& Knobloch]{ecke1992hopf}
{\sc \au{Ecke, R.~E.}, \au{Zhong, F.} \& \au{Knobloch, E.}} \yr{1992}  \at{Hopf bifurcation with broken reflection symmetry in rotating rayleigh-b{\'e}nard convection}.  \jt{Europhysics Letters}  \bvol{19}~(3),  \pg{177}.

\bibitem[Favier \& Knobloch(2020)]{favier2020robust}
{\sc \au{Favier, B.} \& \au{Knobloch, E.}} \yr{2020}  \at{Robust wall states in rapidly rotating {Rayleigh--B{\'e}nard} convection}.  \jt{J. Fluid Mech.}  \bvol{895},  \pg{R1}.

\bibitem[Herrmann \& Busse(1993)]{herrmann_busse_1993}
{\sc \au{Herrmann, J.} \& \au{Busse, F.~H.}} \yr{1993}  \at{Asymptotic theory of wall-attached convection in a rotating fluid layer}.  \jt{J.~Fluid Mech.}  \bvol{255},  \pg{183–194}.

\bibitem[Houchens {\em et~al.\/}(2002)Houchens, Witkowski \& Walker]{houchens2002rayleigh}
{\sc \au{Houchens, B.~C.}, \au{Witkowski, L.~M.} \& \au{Walker, J.~S.}} \yr{2002}  \at{{Rayleigh--B{\'e}nard} instability in a vertical cylinder with a vertical magnetic field}.  \jt{J. Fluid Mech.}  \bvol{469},  \pg{189--207}.

\bibitem[Jones(2011)]{Jones2011}
{\sc \au{Jones, C.~A.}} \yr{2011}  \at{{Planetary magnetic fields and fluid dynamos}}.  \jt{Annu. Rev. Fluid Mech.}  \bvol{43},  \pg{583--614}.

\bibitem[Liao {\em et~al.\/}(2006)Liao, Zhang \& Chang]{liao_zhang_chang_2006}
{\sc \au{Liao, X.}, \au{Zhang, K.} \& \au{Chang, Y.}} \yr{2006}  \at{On boundary-layer convection in a rotating fluid layer}.  \jt{J.~Fluid Mech.}  \bvol{549},  \pg{375–384}.

\bibitem[Liu {\em et~al.\/}(2018)Liu, Krasnov \& Schumacher]{Liu2018}
{\sc \au{Liu, W.}, \au{Krasnov, D.} \& \au{Schumacher, J.}} \yr{2018}  \at{Wall modes in magnetoconvection at high {H}artmann numbers}.  \jt{J. Fluid Mech.}  \bvol{849},  \pg{R21--R212}.

\bibitem[Ni \& Li(2012)]{ni2012consistent}
{\sc \au{Ni, M.~J.} \& \au{Li, J.~F.}} \yr{2012}  \at{A consistent and conservative scheme for incompressible {MHD} flows at a low magnetic {R}eynolds number. {Part III}: {O}n a staggered mesh}.  \jt{J. Comp. Phys.}  \bvol{231}~(2),  \pg{281--298}.

\bibitem[Reiter {\em et~al.\/}(2022)Reiter, Zhang \& Shishkina]{reiter2022flow}
{\sc \au{Reiter, P.}, \au{Zhang, X.} \& \au{Shishkina, O.}} \yr{2022}  \at{Flow states and heat transport in {Rayleigh--B{\'e}nard} convection with different sidewall boundary conditions}.  \jt{J. Fluid Mech.}  \bvol{936},  \pg{A32}.

\bibitem[Schumacher(2022)]{schumacher2022various}
{\sc \au{Schumacher, J.}} \yr{2022}  \at{The various facets of liquid metal convection}.  \jt{J. Fluid Mech.}  \bvol{946},  \pg{F1}.

\bibitem[Shilnikov(1965)]{shilnikov1965case}
{\sc \au{Shilnikov, L.~P.}} \yr{1965} A case of the existence of a denumerable set of periodic motions.  \bt{In {\em Doklady Akademii Nauk\/}}, ,  \vol{vol. 160},  \pg{pp. 558--561}. Russian Academy of Sciences.

\bibitem[Teimurazov {\em et~al.\/}(2023)Teimurazov, McCormack, Linkmann \& Shishkina]{teimurazov2023unifying}
{\sc \au{Teimurazov, A.}, \au{McCormack, M.}, \au{Linkmann, M.} \& \au{Shishkina, O.}} \yr{2023}  \at{Unifying heat transport model for the transition between buoyancy-dominated and {L}orentz-force-dominated regimes in quasistatic magnetoconvection}.  \jt{arXiv preprint arXiv:2308.01748} .

\bibitem[Wiggins(1988)]{wiggins2013global}
{\sc \au{Wiggins, S.}} \yr{1988} {\em Global bifurcations and chaos: analytical methods\/}.  \publ{Springer}.

\bibitem[Xu {\em et~al.\/}(2023)Xu, Horn \& Aurnou]{xu2023transition}
{\sc \au{Xu, Y.}, \au{Horn, S.} \& \au{Aurnou, J.~M.}} \yr{2023}  \at{The transition from wall modes to multimodality in liquid gallium magnetoconvection}.  \jt{arXiv preprint arXiv:2303.08966} .

\bibitem[Zhang \& Liao(2009)]{zhang_liao_2009}
{\sc \au{Zhang, K.} \& \au{Liao, X.}} \yr{2009}  \at{The onset of convection in rotating circular cylinders with experimental boundary conditions}.  \jt{J.~Fluid Mech.}  \bvol{622},  \pg{63–73}.

\bibitem[Z{\"u}rner {\em et~al.\/}(2020)Z{\"u}rner, Schindler, Vogt, Eckert \& Schumacher]{zurner2020flow}
{\sc \au{Z{\"u}rner, T.}, \au{Schindler, F.}, \au{Vogt, T.}, \au{Eckert, S.} \& \au{Schumacher, J.}} \yr{2020}  \at{Flow regimes of {Rayleigh--B{\'e}nard} convection in a vertical magnetic field}.  \jt{J. Fluid Mech.}  \bvol{894},  \pg{A21}.

\end{thebibliography}

\end{document}